\documentclass[conference]{IEEEtran}
\usepackage[english]{babel}
\usepackage[T1]{fontenc} 
\usepackage{verbatim} 
\usepackage{fancyhdr, ifpdf}
\usepackage{tabularx} 
\usepackage{listings} 
\usepackage{graphicx}
\usepackage[figure,boxed,ruled,vlined]{algorithm2e}
\usepackage{algpseudocode}
\usepackage{rotating}
\usepackage{attachfile}
\usepackage{balance}
\usepackage{amsmath}
\usepackage{textpos}
\usepackage{numprint}
\usepackage{paralist}
\npthousandsep{,}

\usepackage{listings}
\usepackage{framed}

\usepackage[table]{xcolor}

\newcommand\companiondoc{the companion technical report \cite{paperdata}}

\usepackage{xcomment}

\title{Mining Software Repair Models for Reasoning on the Search Space of Automated Program Fixing}

\author{\IEEEauthorblockN{Matias Martinez}
\IEEEauthorblockA{University of Lille \& INRIA}
\and
\IEEEauthorblockN{Martin Monperrus}
\IEEEauthorblockA{University of Lille \& INRIA}}

\newlength{\figureskip}
 
\begin{document}

\maketitle

\begin{textblock*}{\textwidth}(0cm,-1.3cm)
\begin{center}
Empirical Software Engineering, Springer, 2013 (accepted for publication on Sep. 11, 2013). 
\end{center}
\end{textblock*}

\begin{abstract}
This paper is about understanding the nature of bug fixing by analyzing thousands of bug fix transactions of software repositories. It then places this learned knowledge in the context of automated program repair. 
We give extensive empirical results on the nature of human bug fixes at a large scale and a fine granularity with abstract syntax tree differencing.
We set up mathematical reasoning on the search space of automated repair and the time to navigate through it.
By applying our method on 14 repositories of Java software and \numprint{89993} versioning transactions,  we show that not all probabilistic repair models are equivalent.
\end{abstract}

\section{Introduction}

Automated program fixing consists of generating source code in order to fix bugs in an automated manner \cite{weimer2006patches,Weimer2009,Wei2010,Dallmeier2009,Arcuri20113494}.
The generated fix is often an incremental modification (a ``patch'' or  ``diff'') over the software version exhibiting the bug.
The previous contributions in this new research field make different assumptions on what is required as input (e.g. good test suites \cite{Weimer2009}, pre- and post-conditions \cite{Wei2010}, policy models \cite{weimer2006patches}). 
The repair strategies also vary significantly. Examples of radically different models include genetic algorithms \cite{Weimer2009} and satisfiability models (SAT) \cite{Gopinath}.

In this paper, we take a step back and look at the problem from an empirical perspective. What are real bug fixes made of? 
The kind of results we extensively discuss later are for instance: in bug-fixes of open source software projects, the most common source code change consists of inserting a method invocation. 
Can we reuse the knowledge for reasoning on automated program repair? We propose a framework to do so, by reasoning on the kind of bug fixes. 
This framework enables us to show that the granularity of the analysis of real commits (which we call ``repair models'') has a big impact on the navigation into the search space of program repair. 
We further show that the heuristics used to build probability distributions on top the repair models also make a significant difference: not all repair actions are equals!

Let us now make precise what we mean with  \emph{repair actions} and  \emph{repair models}.
A software repair action is a kind of modification on source code that is made to fix bugs.
We can cite as examples: 
changing the initialization of a variable;
adding a condition in an ``if'' statement;
adding a method call, etc.
In this paper, we use the term ``repair model'' to refer to a set of repair actions.
For instance, the repair model of Weimer et al. \cite{Weimer2009} has three repair actions: deleting a statement, inserting a statement taken from another part of the software, swapping two statements

There is a key difference between a repair action and a repair: a repair action is a kind of repair, a repair is a concrete patch. In object-oriented terminology, a repair is an instance of a repair action.  For instance, ``adding a method call'' is a repair action, ``adding \texttt{x.foo()}'' is a repair.
A repair action is program- and domain-independent, it contains no domain-specific data such as variable names or literal values.

First we present an approach to mine repair actions from patches written by developers. 
We find traces of human-based program fixing in software repositories (e.g. CVS, SVN or Git), where there are versioning transactions (a.k.a commits) that only fix bugs.
We use those ``fix transactions'' to mine AST-level repair actions such as adding a method call, changing the condition of a ``if'', deleting a catch block.
Repair actions are extracted with the abstract differencing algorithm of Fluri et al. \cite{Fluri2007b}.  
This results in repair models that are much bigger (41 and 173 repair actions) compared to  related work which considers at most a handful of repair actions. 

Second, we propose to decorate the repair models with a probability distribution.
Our intuition is that not all repair actions are equal and certain repair actions are more likely to fix bugs than others. 
We also take an empirical viewpoint to define those probability distributions: we learn them from software repositories.
We show that those probability distributions are independent of the application domain.

Third, we demonstrate that our probabilistic repair models enable us to reason on the search space of automated program repair.
The multinomial theorem \cite[p.73]{bona2011} comes into play to analyze the time to navigate into the search space of automated repair from a theoretical viewpoint.

To sum up, our contributions are:
\begin{itemize}
\item An extensive analysis of the content of software versioning transactions: our analysis is novel both with respect of size (\numprint{89,993} transactions of 14 open-source Java projects) and granularity (173 repair actions at the level of the AST). 
\item A probabilistic mathematical reasoning on automated repair showing that depending on the viewpoint one may quickly navigate -- or not -- into the search space of automated repair. Despite being theoretical, our results highlight an important property of the deep structure of this search space: the likely-correct repairs are highly concentrated in some parts of the search space, as stars are concentrated into galaxies in our universe.
\end{itemize}

This article is a revised version of a technical report \cite{Martinez2012}. 
It reads as follows.
Section \ref{sec:change-spaces} describes how we map concrete versioning transactions to change actions.
Section \ref{sec:transaction-bags} discusses how to only select bug fix transactions.
Section \ref{sec:repair-actions} then shows that those change actions are actually repair actions under certain assumptions.
Section \ref{sec:bug-repair-simulation} presents our theoretical analysis on the time to navigate in the search space of automated repair.
Finally, we compare our results with the related work (in Section \ref{sec:relatedwork}) and concludes.

\section{Describing Versioning Transactions with a Change Model}
\label{sec:change-spaces}

In this section, we describe the contents of versioning transactions of 14 repositories of Java software.
Previous empirical studies on versioning transactions \cite{Hindle2008,German2006,Hattori2008,Alali2008,Purushothaman2005} focus on metadata (e.g., authorship, commit text) or size metrics (number of changed files, number of hunks, etc.).
On the contrary, we aim at describing versioning transactions in terms of contents: what kind of source code change they contain: addition of method calls; modification of conditional statements; etc.
There is previous work on the evolution of source code (e.g. \cite{Livshits2005,Robbes2008,Giger2011}).
However, to our knowledge,  they are all at a coarser granularity compared to what we describe in this paper.

Note that other terms exist for referring to versioning transactions: ``commits'', ``changesets'', ``revisions''. Those terms reflect the competition between versioning tools (e.g. Git uses ``changeset'' while SVN ``revision'') and the difference between technical documentation and academic publications which often use ``transaction''. In this paper, we equate those terms and generally use the term ``transaction'', as previous research does.

Software versioning repositories (managed by version control systems such as CVS, SVN or Git) store the source code changes made by developers during the software lifecycle. 
Version control systems (VCS) enables developers to query versioning transactions based on revision number, authorship, etc. For a given transaction, VCS can produce a difference (``diff'') view that is a line-based difference view of source code. 
For instance, let us consider the following diff:
\begin{lstlisting}
 while(i < MAX_VALUE){
  op.createPanel(i);
- i=i+1;
+ i=i+2;
 }
\end{lstlisting}
The difference shows one line replaced by another one. 
However, one could also observe the changes at the abstract syntax tree (AST) level, rather than at the line level.
In this case, the AST diff is an update of an assignment statement within a for loop. 
In this section, our research question is: \emph{what are versioning transactions made of at the abstract syntax tree level?}.

To answer this question, we have followed the following methodology.
First, we have chosen an AST differencing algorithm from the literature.
Then, we have constituted a dataset of software repositories to run the AST differencing algorithm on a large number of transactions.
Finally, we have computed descriptive statistics on those AST-based differences.
Let us first discuss the dataset.

\subsection{Dataset}\label{cap:dataset}

CVS-Vintage is a dataset of 14 repositories of open-source Java software \cite{Monperrus2012toappear}.
The inclusion criterion of CVS-Vintage is that the repository mostly contains Java code and has been used in previous published academic work on mining software repositories and software evolution.
This dataset covers different domains: desktop applications, server applications, libraries such as logging, compilation, etc. 
It includes the repositories of the following projects:
ArgoUML, Columba, JBoss, JHotdraw, Log4j, org.eclipse.ui.workbench, Struts,
Carol, Dnsjava, Jedit, Junit, org.\-eclipse\-.jdt\-.core, Scarab and Tomcat.
In all, the dataset contains \numprint{89993} versioning transactions, \numprint{62179} of them have at least one modified Java file. Overtime, \numprint{259264} Java files have been revised (which makes a mean number of 4.2 Java files modified per transaction).

\subsection{Abstract Syntax Tree Differencing}

There are different propositions of AST differencing algorithms in the literature.
Important ones include  Raghavan et al.'s Dex \cite{Raghavan2004}, Neamtiu et al's AST matcher \cite{Neamtiu2005}
and Fluri et al's ChangeDistiller \cite{Fluri2007b}.
For our empirical study on the contents of versioning transactions, we have selected the latter. 

ChangeDistiller \cite{Fluri2007b} is a fine-grain AST differencing tool for Java. 
It expresses fine granularity source code changes using a taxonomy of 41  source changes types, such as ``statement insertion'' of ``if conditional change''.
ChangeDistiller handles changes that are specific to object-oriented elements such as ``field addition''. 
Fluri and colleagues have published an open-source stable and reusable implementation of their algorithm for analyzing AST changes of Java code.

Change\-Distiller produces a set of ``source code changes'' for each pair of Java files from versioning transactions.
For a source code change, the main output of Change\-Distiller is a ``change type''  (from the taxonomy aforementioned).
However, for our analysis, we also consider two other pieces of information.
We reformulate the output of Change\-Distiller, each AST source code change is represented as a 2-value tuple:
$scc = (ct, et)$
where \emph{ct} is one of the 41 change types, \emph{et} (for entity type) refers to the source code entity related to the change (for instance, a statement update may change a method call or an assignment).
Since ChangeDistiller is an AST differencer, formatting transactions (such as changing the indentation) produce no AST-level change at all.
The short listing above would be represented as one single AST change that is a statement update (ct) of an assignment (et).

 \subsection{Change Models}
\label{sec:change-models}
 
All versioning transactions can be expressed within a ``change model''.
We define a change model as a set of ``change actions''. 
For instance, the change model of standard Unix diff is composed of two change actions: line addition and line deletion.
A change model represents a kind of feature space, and observations in that space can be valued. For instance, a standard Unix diff produces two integer values: the number of added lines and the number of deleted lines.
ChangeDistiller enables us to define the following change models.

\textbf{CT} (\emph{Change Type}) is composed of 41 features, the 41 change types of ChangeDistiller.
For instance, one of this feature is ``Statement Insertion'' (we may use the shortened name ``Stmt\_Insert'').
\textbf{CTET} (\emph{Change Type Entity Type}) is made of all valid combinations of the Cartesian product between change types and entity types.
CTET is a refinement of CT. Each repair action of CT is mapped to $[1\ldots n]$ repair actions of CTET. Hence the labels of the repair actions of CTET always contain the label of CT.
There are 104 entity types and 41 change types but many combinations are impossible by construction, as a result  CTET contains 173 features.
For instance, since there is one entity type representing assignments, one feature of CTET is ``statement insertion of an assignment''. 

In the rest of this paper, we express versioning transactions within those two change models. 
There is no better change model per se: they describe versioning transactions at different granularity.
We will see later that, depending on the perspective, both change models have pros and cons.

 \subsection{Measures for Change Actions}

 We define two measures for a change action $i$:
 $\alpha_i$ is the absolute number of change action $i$ in a dataset;
$\chi_i$ is the probability of observing a change action $i$ as given by its frequency over all changes ($\chi_i= \alpha_i/\sum \alpha_i$).
For instance, let us consider feature space $CT$ and the change action ``statement insertion'' (StmtIns).
If there is $\alpha_{StmtIns} =  12$ source code changes  related to statement insertion among 100,
the probability of observing a statement insertion is $\chi_{StmtIns} = 12\%$.

\subsection{Empirical Results}

We have run ChangeDistiller over the \numprint{62179} Java transactions of our dataset, resulting in   \numprint{1196385} AST-level changes for both change models.
For change model CT, which is rather coarse-granularity, the three most common changes are ``statement insert'' (28\% of all changes), ``statement delete'' (23\% of all changes) and ``statement update'' (14\% of all changes). Certain changes are rare, for instance, ``addition of class derivability'' (adding keyword \texttt{final} to the class declaration) only appears 99 times (0.0008\% of all changes). The complete results are given in \companiondoc.

Table  \ref{tab:top10ctet}  presents the top 20 change actions and the associated measures for change model CTET.
The comprehensive table for all 173 change actions is given in \companiondoc.
In Table \ref{tab:top10ctet}, one sees that inserting method invocations as statement is the most common change, which makes sense for open-source object-oriented software that is growing.

\begin{table}
\scriptsize
 \rowcolors[]{2}{gray!10}{}
\begin{center}
 \begin{tabularx}{\columnwidth}{ p{5.cm} | X | X }
                                             Change Action &                                                   $\alpha_i$ &                                                     Prob. $\chi_i$ \\
\hline
 Statement insert of method invocation &                                                               \numprint{83046} &                                                                 6.9\% \\
                               Statement insert of if statement &                                                                \numprint{79166} &                                                                 6.6\% \\
                          Statement update of method invocation &                                                               \numprint{76023}  &                                                                 6.4\% \\
                          Statement delete of method invocation &                                                               \numprint{65357}  &                                                                 5.5\% \\
                               Statement delete of if statement &                                                                \numprint{59336}  &                                                                   5\% \\
            Statement insert of variable declaration statement &                                                                \numprint{54951}  &                                                                 4.6\% \\
                                  Statement insert of assignment &                                                                \numprint{49222}  &                                                                 4.1\% \\
                              Additional functionality of method &                                                               \numprint{49192}  &                                                                 4.1\% \\
            Statement delete of variable declaration statement &                                                              \numprint{44519}  &                                                                 3.7\% \\
            Statement update of variable declaration statement &                                                        \numprint{41838}  &                                                                 3.5\% \\
                                  Statement delete of assignment &                                                      \numprint{41281}  &                                                                 3.5\% \\
                   Condition expression change of if statement &                                                    \numprint{40415}  &                                                                 3.4\% \\
                                  Statement update of assignment &                                                     \numprint{34802}  &                                                                 2.9\% \\
                           Addition of attribute &                                                         \numprint{29328}  &                                                                 2.5\% \\
                                 Removal of method &                                                   \numprint{26172}  &                                                                 2.2\% \\
                           Statement insert of return statement &                                               \numprint{24184}  &                                                                   2\% \\
                  Statement parent change of method invocation &                                       \numprint{21010}  &                                                                 1.8\% \\
                           Statement delete of return statement &                                            \numprint{20880}  &                                                                 1.7\% \\
                     Insert of else statement &                                            \numprint{20227}  &                                                                 1.7\% \\
                     Deletion of else statement &                                            \numprint{17197}  &                                                                 1.4\% \\             

  {\bf Total} &                                   {\bf   \numprint{1196385}} &                                        \\
  \end{tabularx}
  \caption{The abundance of AST-level changes of change model CTET over 62,179 versioning Transactions. The probability $\chi_i$ is the relative frequency over all changes (e.g. 6.9\% of source code changes are insertions of method invocation).}
  \label{tab:top10ctet}
\end{center}
\vspace{\figureskip}
  \end{table}

Let us now compare the results over change models CT and CTET. One can see that statement insertion is mostly composed of inserting a method invocation (6.9\%), insert an ``if'' conditionals (6.6\%), and insert a new variable (4.6\%).
Since change model CTET is at a finer granularity, there are less observations: both $\alpha_i$ and $\chi_i$ are lower. 
The probability distribution ($\chi_i$) over the change model is less sharp (smaller values) since the feature space is bigger.
High value of  $\chi_i$ means that we have a change action that can frequently be found in real data: those change actions have of a high ``coverage'' of data.
CTET features describe modifications of software at a finer granularity.
\emph{The differences between those two change models illustrate the tension between a high coverage and the analysis granularity.}

\subsection{Project-independence of Change Models}

An important question is whether the probability distribution (composed of all $\chi_i$) of Table \ref{tab:top10ctet} is generalizable to Java software or not.
That is, do developers evolve software in a similar manner over different projects?
To answer this question, we have computed the metric values not for the whole dataset, but per project.
In other words, we have computed the frequency of change actions in 14 software repositories.
We would like to see that the values do not vary between projects, which would mean that the probability distributions over change actions are project-independent. Since our dataset covers many different domains, having high correlation values would be a strong point towards generalization.

As correlation metric, we use Spearman's $\rho$. We choose  Spearman's $\rho$ because it is non-parametric. 
In our case, what matters is to know whether the importance of change actions is similar (for instance that ``statement update'' is more common than``condition expression change''). Contrary to parametric correlation metric (e.g. Pearson), Spearman's $\rho$ only focuses on the ordering between change actions, which is what we are interested in.

We compute the Spearman correlation values between the probability distributions of all pairs of project of our datasets (i.e. $\frac{14*13}{2}=91$ combinations). One correlation value takes as input two vectors representing the probability distributions (of size 41 for change model CT and 173 for change model CTET).

The critical value of Spearman's $\rho$ depends on size of the vectors being compared and on the required confidence level.
At confidence level $\alpha=0.01$, the critical value for change model CT (41 features) is 0.364 and is 0.301\footnote{Most statistical tables of Spearman's $\rho$ stop at N=60, however since the critical values decreases with N, if $\rho>0.301$ the null hypothesis is still rejected.} for change model CTET   (values from statistical tables, we used  \cite{stat-tables}). If the correlation is higher than the critical value, the null hypothesis (a random distribution) is rejected.

\begin{figure}
\includegraphics[width=\columnwidth]{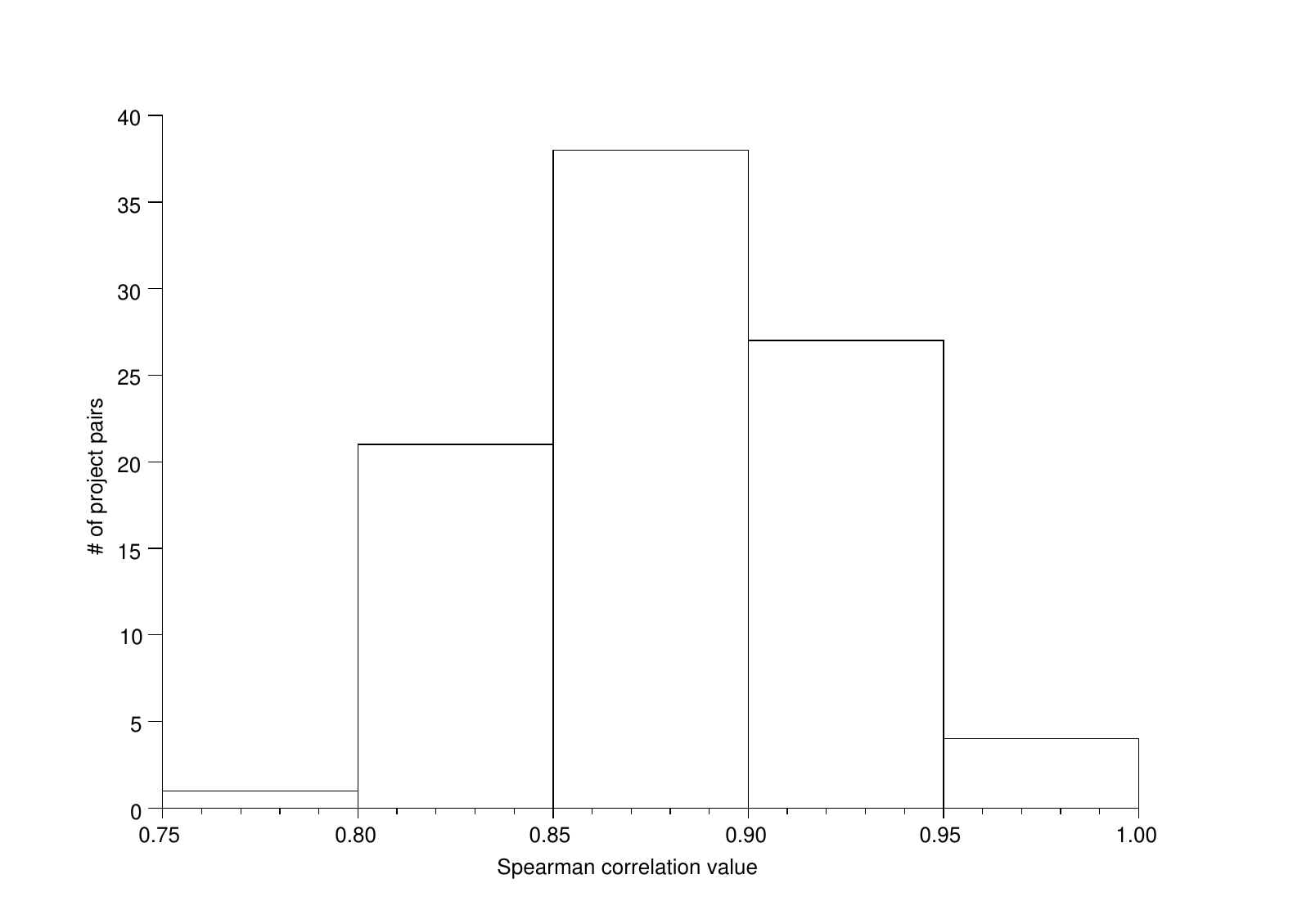}
\caption{Histogram of the Spearman Correlation between Changes Action Frequencies of Change Model CT Mined on Different Projects. There is no outlier, there are all higher than 0.75, meaning that the importance of change actions is project-independent.}
\label{fig:fig-correlation-histogram}
\end{figure}

For instance, in change model CT, the Spearman correlation between Columba and ArgoUML is 0.94 which is much higher than the critical value (0.364). This means that the correlation is statistically significant at $\alpha=0.01$ confidence level. The high value shows that those two projects were evolved in a very similar manner. 
All values are given in \companiondoc.
Figure \ref{fig:fig-correlation-histogram} gives the distribution of Spearman correlation values for change model CT. 
75\% of the pairs of projects have a Spearman correlation higher than 0.85\footnote{
Spearman correlation is based on ranks, a value of 0.85 means either that most change actions are ranked similarly or that a single change action has a really different rank.}.
For all pairs of projects, in change model CT, Spearman's $\rho$ is much higher that the critical value.
\emph{This shows that the likelihood of observing a change action is globally independent of the project used for computing it.}

To understand the meaning of those correlation values, let us now analyze in detail the lowest and highest correlation values.
The highest correlation value is 0.98 and it corresponds to the project pair Eclipse-Workbench and Log4j. In this case, 33 out of 41 change actions have a rank difference between 0 and 3. 
The lowest correlation value is 0.80 and it corresponds to Spearman correlation values between projects Tomcat and Carol. 
In this case, the maximum rank change is 23 (for change action ``Removing Method Overridability'' --- removing \emph{final} for methods). In total, between Tomcat and Carol, there are six change actions for which the importance changes of at least 10 ranks. Those high values trigger the 0.80 Spearman correlation. However, for common changes, it turns out that their ranks do not change at all (e.g. for ``Statement Insert'', ``Statement Update'', etc.). 

We have also computed the correlation between projects within change model CTET (see \companiondoc). 
They are all above 0.301, the critical value for vectors of size 173 at $\alpha=0.01$ confidence level, showing that in change model CTET, the change action importance is project-independent as well, in a statistically significant manner.
Despite being high, we note that they are slightly lower than for change model CT, this is due to the fact that Spearman's $\rho$ generally decreases with the vector size (as shown by the statistical table).

\subsection{Recapitulation} 

To sum up,
we provide the empirical importance of 173 source code change actions;
we show that the importance of change actions are project independent;
we show that the probability distribution of change actions is very unbalanced.

Our results are based on the analysis of   \numprint{62179}  transactions. To our knowledge, those results have never been published before, given this analysis granularity and the scale of the empirical study.

The threats to the validity of our results are of two kinds.
From the internal validity viewpoint, a bug somewhere in the implementation may invalidate our results.
From the external validity viewpoint, there is risk that our dataset of 14 projects is not representative of Java software as a whole, even if they are written by different persons from different organizations in different application domains. Also, our results may not generalize to other programming languages.

\section{Slicing Transactions to Focus on Bug Fixes}
\label{sec:transaction-bags}

In Section \ref{sec:change-spaces}, we have defined and discussed two measures per change action $i$: $\alpha_i$ and $\chi_i$.
For instance, $\chi_{StmtInsert}$ gives the frequency of a statement insertion.
Those measures implicitly depend on a transaction bag to be computed.
So far we have considered all versioning transactions of the repository.
For defining a repair space, we need to apply those two measures on a transaction bag representative of software repair.
How should we slice transactions to focus on bug fixes?

An intuitive method, that we will use as baseline, is to rely on the commit message (by slicing only those transactions that contain a given word or expression related to bug fixing). Before going further, let us clarify the goal of the classification: 
the goal is to have a good approximation of the probability distribution of change actions for software repair\footnote{Note that our goal is not to have a good classification in terms of precision or recall.}.
Later is the paper, we will define a mathematical criterion to tell whether one approximation is better than another.

\subsection{Slicing Based on the Commit Message}

When committing source code changes, developers may write a comment/message explaining the changes they have made.  For instance when a transaction is related to a bug fix, they may write a comment referencing the bug report or describing the fix.

To identify transaction bags related to bug fix, previous work focused on the content of the commit text: whether it contains a bug identifier, or whether it contains some keywords such as ``fix'' (see \cite{Murgia2010} for a discussion on those approaches).
To identify bug fix patterns, Pan et al. \cite{Pan2008} select transactions containing at least one occurrence of ``bug'', ``fix'' or ``patch''.
We call this transaction bag BFP.
We will compute $\alpha_i$ and $\chi_i$ based on this definition.

Such a transaction bag makes a strong assumption on the development process and the developer's behavior: it assumes that developers generally put syntactic features in commit texts enabling to recognize repair transactions, which is not really true in practice \cite{Murgia2010,Wu2011,Bird2009}.

\subsection{Slicing  Based on the Change Size in Terms of Number of AST Changes}
We may also define fixing transaction bags based on their ``AST diffs'', i.e.; based on the type and numbers of change actions that a versioning transaction contains.
This transaction bag is called N-SC (for N Abstract \emph{S}yntactic Changes), e.g. 5-SC represents the bag of transactions containing five AST-level source code change.

In particular, we assume that small transactions are very likely to only contain a bug  fix and unlikely to contain a new feature.
Repair actions may be those that appear atomically in transactions (i.e. the transaction only contains one AST-level source code change).
``1-SC'' (composed of all transactions of one single AST change) is the transaction bag that embodies this assumption.
Let us verify this assumption.

\subsection{Do Small Versioning Transactions Fix Bugs?}

\subsubsection{Experiment}
We set up a study to determine whether small transactions correspond to bug fixes changes.
We define small as  those transactions that introduce only one AST change. 
 
\subsubsection{Overview}\label{sec:bfteval}
The study consists in manual inspection and evaluation of source code changes of versioning transactions.
First, we randomly take a sample set of transactions from our dataset (see \ref{cap:dataset}).
Then, we create an ``evaluation item'' for each pair of files of the sample set (the file before and after the revision). 
An evaluation item contains data to help the raters to decide whether a transaction is a bug fix or not:
the syntactic line-based differencing between the revision pair of the transaction (it helps to visualize the changes), the AST change between them (type and location -- e.g. insertion of method invocation at line 42) and the commit message associated with the transaction.

\subsubsection{Sampling Versioning Transactions}
 
We use stratified sampling to randomly select  1-SC versioning transactions from the software history of 16 open source projects (mostly from \cite{Monperrus2012toappear}). 
Recall that a ``1-SC'' versioning transaction only introduces one AST change.
The stratification consists of picking 10 items (if 10 are found) per project.
In total, the sample set contains 144 transactions sampled over \numprint{6953} 1-SC transactions present in our dataset.  

\subsubsection{Evaluation Procedure}
The 144 evaluation items were evaluated by three people called the \emph{raters}: the paper authors and a colleague, member of the faculty at the University of Bordeaux.
During the evaluation, each item (see \ref{sec:bfteval}) is presented to a rater, one by one. 
The rater has to answer the question \emph{Is a bug fix change?}. 
The possible answers are
\begin{inparaenum}[\itshape a\upshape)]
\item \emph{Yes, the change is a bug fix},  
\item \emph{No, the change is not a bug fix} and 
\item \emph{I don't know}. 
\end{inparaenum}
Optionally, the rater can write a comment to explain his decision.

\subsubsection{Experiment Results}

\paragraph{Level of Agreement}
The three raters fully agreed that 74 of 144 (51.8\%) transactions from the sample transactions are bug fixes. 
If we consider the majority (at least 2/3 agree), 95 of 144 transactions (66\%) were considered as bug fix transactions.
The complete rating data is given in \companiondoc.
 
Table  \ref{tab:agreement} presents the number of agreements.  
The column \emph{Full Agreement} shows the number of transactions for which all raters agreed. 
For example, the three rates agreed that there is a bug fix in 74/144 transactions. 
The \emph{Majority} column shows the number of transactions for which two out of three raters agree. 
To sum up, small transactions predominantly consists of bug fixes.

\begin{table}[]
\begin{center}
  \begin{tabular}{rrrr}
 &	Full Agreement (3/3)&	Majority	(2/3) \\ 
  \hline
Transaction is a Bug Fix& 	74& 	21	\\ \hline
Transaction is not a Bug Fix& 	22& 	23\\ \hline 
I don't know& 	0	& 1	\\ \hline
  \end{tabular}
\end{center}
\caption{The Results of The Manual Inspection of 144 Transactions by Three Raters.
}
\label{tab:agreement}
\end{table}

Among the transactions with full agreement on the absence of bug fix changes, the most common case found was the addition of a method. This change indeed consists of the addition of one single AST change (the addition of a ``method'' node).
Interestingly, in some cases, adding a method was indeed a bug fix, when polymorphism is used: the new method fixes the bug by replacing the super implementation.

\paragraph{Statistics}
Let us assume that $p_i$ measures the degree of agreement for a single item (in our case in $\{\frac{1}{3},\frac{2}{3},\frac{3}{3}\}$.
The overall agreement $\bar{P}$ \cite{cohen1960} is the average over $p_i$.  We have $\bar{P} = 0.77$. Using the scale introduced by \cite{landis1977}, this value means there is a \emph{Substantial} overall agreement between the rates, close to an \emph{Almost perfect agreement}. 

The coefficient $\kappa$ (Kappa) \cite{cohen1960,fleiss1971} measures the confidence in the agreement level by removing the chance factor\footnote{Some degree of agreement is expected when the ratings are purely random\cite{cohen1960,fleiss1971}.}. 
The $\kappa$ degree of agreement in our study is 0.517, a value distant from the critical value (it is 0).
The null hypothesis is rejected, the observed agreement was not due to chance.

\subsubsection{Conclusion}
The manual inspection of  144 versioning transaction shows that there is a relation between the one AST change transactions and bug fixing.
Consequently, we can use the 1-SC transaction bag to estimate the probability of change actions for software repair.

\section{From Change Models to Repair Models}
\label{sec:repair-actions}

This section presents how we can transform a ``change model'' into a ``repair model'' usable for automated software repair.
As discussed in Section \ref{sec:change-spaces}, a change model describes all types of source code change that occur during software evolution.
On the contrary, we define a ``repair action'' as a change action that often occurs for repairing software, i.e. often used for fixing bugs.

By construction,  a repair model is equal to a subset of a change model in terms of features.
But more than the number of features, our intuition is that the probability distribution over the feature space would vary between change models and repair models.
For instance, one might expect that changing the initialization of a variable has a higher probability in a repair model.
Hence, the difference between a change model and a repair model is matter of perspective.
Since we are interested in automated program repair, we now concentrate on the ``repair'' perspective hence use the terms ``repair model'' and ``repair action'' in the rest of the paper.

\begin{table*}[ht]
\begin{center}
\scriptsize
 \rowcolors[]{2}{gray!10}{}
\begin{tabularx}{\textwidth}{X X X X X X}
                       ALL &                        BFP &                       1-SC &                       5-SC &                      10-SC &                      20-SC \\
\hline
         Stmt\_Insert-29\% &          Stmt\_Insert-32\% &             Stmt\_Upd-38\% &          Stmt\_Insert-28\% &          Stmt\_Insert-31\% &          Stmt\_Insert-33\% \\
            Stmt\_Del-23\% &             Stmt\_Del-23\% &            Add\_Funct-14\% &             Stmt\_Upd-24\% &             Stmt\_Upd-19\% &             Stmt\_Del-16\% \\
            Stmt\_Upd-15\% &             Stmt\_Upd-12\% &          Cond\_Change-13\% &             Stmt\_Del-11\% &             Stmt\_Del-14\% &             Stmt\_Upd-16\% \\
         Param\_Change-6\% &          Param\_Change-7\% &          Stmt\_Insert-12\% &            Add\_Funct-10\% &             Add\_Funct-8\% &          Param\_Change-7\% \\
         Order\_Change-5\% &          Order\_Change-6\% &              Stmt\_Del-6\% &           Cond\_Change-7\% &          Param\_Change-7\% &             Add\_Funct-7\% \\
            Add\_Funct-4\% &             Add\_Funct-4\% &             Rem\_Funct-5\% &          Param\_Change-5\% &           Cond\_Change-6\% &           Cond\_Change-5\% \\
          Cond\_Change-4\% &           Cond\_Change-3\% &           Add\_Obj\_St-3\% &           Add\_Obj\_St-3\% &           Add\_Obj\_St-3\% &           Add\_Obj\_St-3\% \\
          Add\_Obj\_St-2\% &           Add\_Obj\_St-2\% &          Order\_Change-2\% &             Rem\_Funct-3\% &             Rem\_Funct-2\% &          Order\_Change-3\% \\
            Rem\_Funct-2\% &      Alt\_Part\_Insert-2\% &           Rem\_Obj\_St-2\% &          Order\_Change-1\% &          Order\_Change-2\% &             Rem\_Funct-2\% \\
     Alt\_Part\_Insert-2\% &             Rem\_Funct-2\% &    Inc\_Access\_Change-1\% &           Rem\_Obj\_St-1\% &      Alt\_Part\_Insert-1\% &      Alt\_Part\_Insert-2\% \\

\end{tabularx}
\caption{Top 10 Change Types of Change Model CT and their Probability $\chi_i$ for Different Transaction Bags. The different heuristics used to compute the fix transactions bags has a significant impact on both the ranking and the probabilities.}
\label{tab:top-10-ct-heuristics}

\end{center}
\vspace{\figureskip}

\end{table*}

\subsection{Methodology}
\label{sec:heuristics}

We have applied the same methodology as in \ref{sec:change-spaces}.
We have computed the probability distributions of repair model CT and CTET based on different definitions of fix transactions, i.e. we have computed $\alpha_i$ and $\chi_i$ based on the transactions bags discussed in \ref{sec:transaction-bags}:
ALL transactions, N-SC and BFP.
For N-SC, we choose four values of N: 1-SC, 5-SC, 10-SC and 20-SC.
Transactions larger than 20-SC have almost the same topology of changes as ALL, as we will show later (see section \ref{sec:repair-models-discussion}).

\emph{The main question we ask is whether those different definitions of ``repair transactions'' yield different topologies for repair models.}

\subsection{Empirical Results}

Table \ref{tab:top-10-ct-heuristics} presents the top 10 change types of repair model CT associated with their probability $\chi_i$ for different versioning transaction bags.
The complete table for all repair actions is given in \companiondoc.
Overall, the distribution of repair actions over real bug fix data is very unbalanced, the probability of observing a single repair action goes from more than 30\% to 0.000x\%. We observe the Pareto effect: the top 10 repair actions account for more than 92\% of the cumulative probability distribution.

Furthermore,  we have made the following observations from the experiment results:

First, the order of repair actions (i.e. their likelihood of contributing to bug repair) varies significantly depending on the transaction bag used for computing the probability distribution. For instance: a statement insertion is \#1 when we consider all transactions (column ALL), but only \#4 when considering transactions with a single AST change (column 1-SC). In this case, the probability of observing a statement insertion varies from 29\% to 12\%.

Second, even when the orders obtained from two different transaction bags resemble such as for ALL and 20-SC, the probability distribution still varies: for instance $\chi_{Stmt\_Upd}$ is 29\% for transaction bag ALL, but jumps to 33\% for transaction bag 20-SC.

Third, the probability distributions for transaction bags ALL and BFP are close: repair action has similar probability values.
As consequence, transaction bag BFP maybe is a random subset of ALL transactions.
All those observations also hold for repair model CTET, the complete table is given in the companion technical report \cite{paperdata}.

\emph{Those results are a first answer to our question: different definitions of ``repair transactions'' yield different probability distribution over a repair model.}

\subsection{Discussion}

We have shown that one can base repair models on different methods to extract repair transaction bags. 
There are certain analytical arguments against or for those different repair space topologies.
For instance, selecting transactions based on the commit text makes  a very strong assumption on the quality of software repository data, but ensures that the selected transactions contain at least one actual repair.
Alternatively, small transactions indicate that they focus on to a single concern, they are likely to be a repair. However, small transactions may only see the tip of the fix iceberg (large transactions may be bug fixing as well), resulting in a distorted probability distribution over the repair space.
At the experimental level, the threats to validity are the same as for Section~\ref{sec:change-spaces}.

\subsubsection{Correlation between Transaction Bags}

\begin{table}[ht]
\begin{center}
\begin{tabular}{r|rrrrr}

       & 1-SC & 5-SC & 10-SC & 20-SC  & BFP\\ 
  \hline
ALL & 0.68 & 0.95 & 0.97 & 0.98  &   0.99 \\ 
\end{tabular}
\end{center}
\caption{The Spearman correlation values between repair actions of transaction bag ``ALL'' and  those from the transaction bags built with 5 different heuristics.}
\label{tab:heuristic-correlation}
\end{table}

To what extent are the 6 transactions bags different?
We have calculated the Spearman correlation values between the probabilities over repairs actions between all pairs of distributions.
In particular, we would like to know whether the heuristics yield significantly different results compared to all transactions (transaction bag ALL). Table \ref{tab:heuristic-correlation} presents these correlation values.

For instance, the Spearman correlation value between ALL and 1-SC is 0.68. 
This value shows, as we have noted before, that there is not a strong correlation between the order of their repair actions of both transaction bags.
In other words, heuristic 1-SC indeed focuses on a specific kind of transactions.

On the contrary,  the value between ALL and BFP is 0.99. 
This means the order of the frequency of repair actions are almost identical.
Moreover, Table \ref{tab:heuristic-correlation} shows the correlation values between N-SC (N = 1, 5, 10 and 20) and ALL tend to 1 (i.e. perfect alignment) when N grows. 
This validates the intuition that the size of transactions (in number of AST changes) is a good predictor to focus on transactions that are different in nature from the normal software evolution.
Crossing this result with the results of our empirical study of 144 -SC transactions, there is some evidence that by concentrating on small transactions, we probably have a good approximation of repair transactions.

\subsubsection{Skewness of Probability Distributions}
\begin{figure}
\includegraphics[scale=0.45]{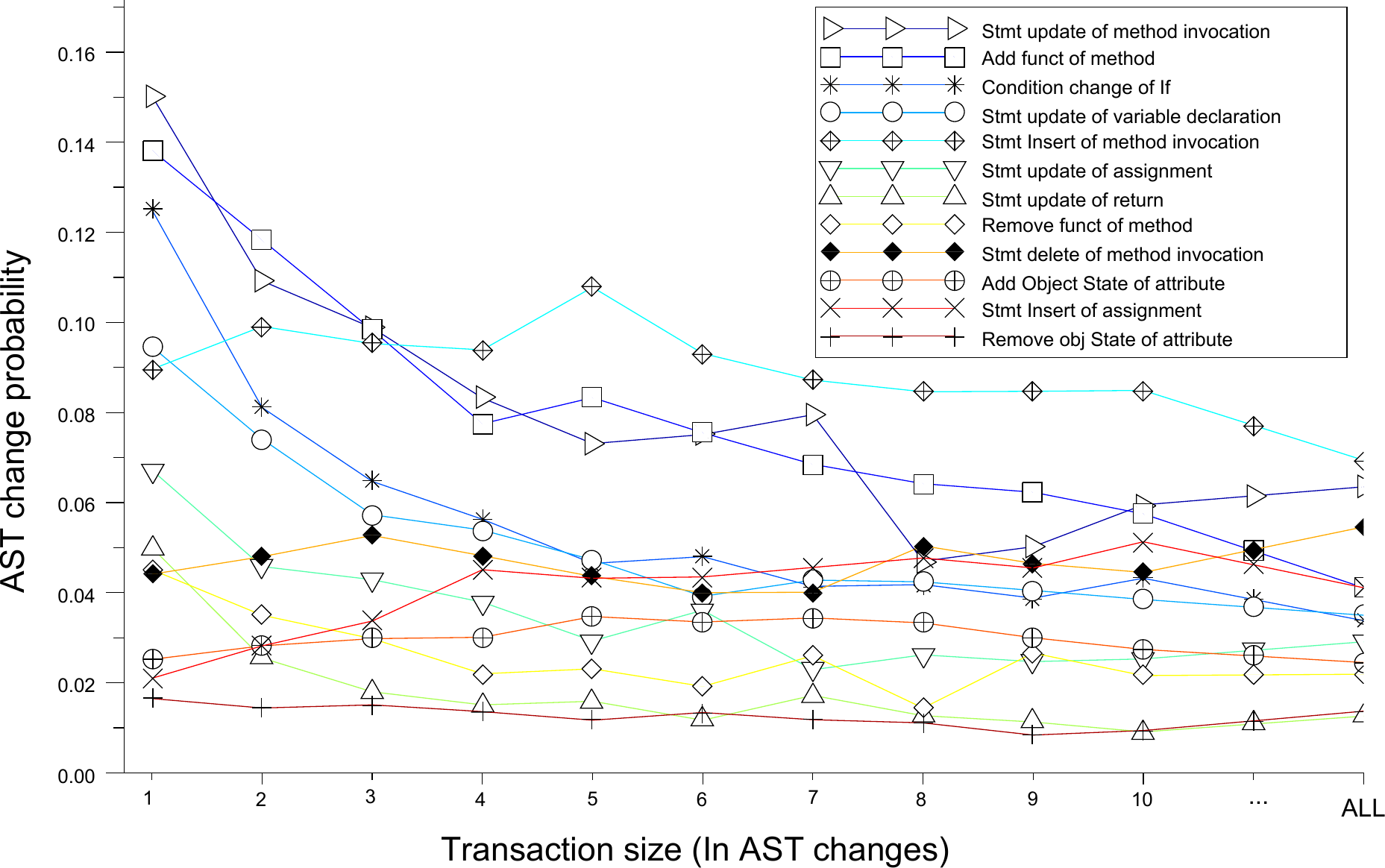}
\caption{Probabilities of the 12 most frequent AST changes for 11 different transaction bags: 10 that include transactions with $i$ AST changes, with $i=1 ... 10$, and the ALL transaction bag.}
\label{fig:fig-ast-change-probability}
\end{figure}

Figure \ref{fig:fig-ast-change-probability} shows the probability for the most frequent repair actions of repair model CTET according to the transaction size (in number of AST changes). 
For instance, the probability of updating a method invocation decreases from 15\% in 1-SC transactions to 7\% in all transactions.
In particular, we observe that:
\begin{inparaenum}[\itshape a\upshape)]
\item for transaction with 1 AST change, the change probabilities are more unbalanced (i.e. less uniform than for all transactions).   
There are 5 changes that are much more frequent than the rest. 
\item for transactions with more than 10 AST changes, the probabilities of top changes are less dispersed and all smaller than 0.9\%
\item the probabilities of those 5 most frequent changes decrease when the transaction size grows.
\end{inparaenum}
This is a further piece of evidence that heuristics N-SC provide a focus on transactions that are of specific nature, different from the bulk of software evolution. 

\label{sec:repair-models-discussion}

\subsubsection{Conclusion}
Those results on repair actions are especially important for automated software repair:
we think it would be fruitful to devise automated repair approaches  that ``imitate'' how human developers fix programs. 
To us, using the probabilistic repair models as described in this section is a first step in that direction.

\section{Automated Analysis of the Time to Navigate into the Search Space of Automated Program Repair}
\label{sec:bug-repair-simulation}

This section discusses the nature of the search space size of automated program repair.
We show that the two repair models defined in \ref{sec:repair-actions} allow mathematical reasoning.
We present a way of comparing repair models and their probability distribution based on data from software repositories. 

\subsection{Decomposing The Repair Search Space}
\label{sec:decomposing-repair}

The search space of automated program repair consists of all explorable bug fixes for a given program and a given bug (whether compilable, executable or correct). If one bounds the size of the repair (e.g. all patched of at most 40 lines), the search space size is finite. 
A naive search space is huge, because even in a bounded size scenario, there are a myriad of elements to be added, removed or modified: statements, variables, operators, literals. 

\emph{A key point of automated program repair research consists of decreasing the time to navigate the repair search space.}

There are many ways to decrease this time. For instance, fault localization enables the search to first focus on places where fixes are likely to be successful.
This one and other components of a repair process may participate in an efficient navigation.
One of them is the ``shaping'' of fixes.

Informally, the shape of a bug fix is a kind of patch. For instance, the repair shape of adding an ``if'' throwing an exception for signaling an incorrect input consists of inserting an if and inserting a throw.
The concept of ``repair shape'' is equivalent to what Wei et al. \cite{Wei2010} call a ``fix schema'', and Weimer et al \cite{Weimer2009} a ``mutation operator''.

In this paper,  we define a ``repair shape'' as an unordered tuple of repair actions (from a set of repair actions called $\mathcal{R}$)\footnote{Since a bug fix may contain several instances of the same repair actions (e.g. several statement insertions), the repair shape may contain several times the same repair action.}.  In the if/throw example aforementioned, in repair space CTET, the repair shape of this bug fix consists of 
two repair actions: statement insertion of ``if'' and  statement insertion of ``throw''. 
The shaping space consists of all possible combinations of repair actions. 

The instantiation of a repair shape is what we call \emph{fix synthesis}.
The complexity of the synthesis depends on the repair actions of the shaping space.  For instance, the repair actions of Weimer et al. \cite{Weimer2009} (insertion, deletion, replace) have an ``easy'' and bounded synthesis space (random picking in the code base). 

To sum up, we consider that the repair search space can be viewed as the combination of 
the fault localization space (where the repair is likely to be successful), 
the shaping space (which kind of repair may be applied)
and the synthesis space (assigning concrete statements and values to the chosen repair actions).
The search space can then be loosely defined as the Cartesian product of those spaces and its size then reads:
$$
|\mbox{\sc Fault Localization}| \times |\mbox{\sc Shape}| \times |\mbox{\sc Synthesis}|
$$

In this paper, we concentrate on the shaping part of the space.
If one can find efficient strategies to navigate through this shaping space, this would contribute to efficiently navigating through the repair search space as a whole, thanks to the combination.

\subsection{Mathematical Analysis Over Repair Models}
\label{sec:MCShaper}

To analyze the shaping space, we now present a mathematical analysis of our probabilistic repair models.
So far, we have two repair models CT and CTET (see \ref{sec:repair-actions}) and different ways to parametrize them.

According to our probabilistic repair model, a good navigation strategy consists on concentrating on likely repairs first: the repair shape is more likely to be composed of frequent repair actions.
That is a repair shape of size $n$ is predicted by drawing $n$ repair actions according to the probability distribution over the repair model.
Under the pessimistic assumption that repair actions are independent\footnote{
Equation (1) holds if and only if we consider them as independent. If they are not, it means that we under-estimate the deep structure of the repair space, hence we over-approximate the time to navigate in the space to find the correct shape. In other words, even if the repair actions are not independent (which is likely for some of them) our conclusions are sound.}, our repair model makes it possible to know the exact median number of attempts $N$ that is needed to find a given repair shape R (demonstration given in \companiondoc): 

\begin{equation}
\label{eq:median-repair-time}
N = k \mbox{ such that } \sum_{i=1}^{k} p(1-p)^{i-1} \geq 0.5
\end{equation}

$$ \mbox{with } p =\frac{n!}{\Pi_{j}{(e_j!)}}  \times \Pi_{r\in R} P_{\mathcal{P}}(r) $$ 
$$ \mbox{where }  e_j \mbox{~is the number of occurrences of } r_j \mbox{~inside } R$$

For instance, the repair of revision 1.2 of Eclipse's CheckedTreeSelectionDialog\footnote{``Fix for 19346 integrating changes from Sebastian Davids'' \url{http://goo.gl/d4OSi}} consists of two inserted statements.
Equation \ref{eq:median-repair-time} tells us that in repair model CT, we would need in average 12 attempts to find the correct repair shape for this real bug. 

Having only a repair shape is far from having a real fix. However, the concept of repair shape associated with the mathematical formula analyzing the time to navigate the repair space is key to compare ways to build a probability distribution over repair models.

\begin{algorithm*}
\KwIn{C\Comment{A bag of transactions}} 
\KwOut{The median number of attempts to find good repair shapes}

\Begin{
   $\Omega \gets \{\}$\Comment{Result set}\\
   $T,E \gets split(C)$ \Comment{Cross-validation: split C into \emph{T}raining and \emph{E}valuation data}\\
   $M \gets train\_model(T)$ \Comment{Train a repair model (e.g. compute a probability distribution over repair actions)}\\
   \For{$s \in E$ \Comment{\em For all repairs observed in the repository}\\} {
      $n \gets compute\_repairability(s, M)$       \Comment{How long to find this repair according to the repair model}\\
      $\Omega \gets R \cup n$  \Comment{Store the ``repairability'' value of $s$}
   }
   \textbf{return} $median(\Omega)$ \Comment{Returning the median number of attempts to find the repair shapes}
}
\caption{An Algorithm to Compare Fix Shaping Strategies. There may be different flavors of functions $split$, $f$ and $computeRepairability$.}\label{bugsim}
\end{algorithm*}

\subsection{Comparing Probability Distributions Over Repair Actions From Versioning History}

\begin{table*}[ht]

\begin{center}
 \rowcolors[]{2}{gray!10}{}
\begin{tabular}{rrrrrrrrr}

                      Repair /Repair Size&                                 1 &                                 2 &                                 3 &                                 4 &                                 5 &                                 6 &                                 7 &                                 8  \\
\hline
                      ArgoUML  &                        {\bf6} (996) &                       {\bf13} (638) &                       {\bf86} (386) &                      {\bf267} (362) &                     {\bf1394} (254) &                     {\bf5977} (234) &                    {\bf16748} (197) &                    {\bf73430} (166) \\
                        Carol  &                         {\bf7} (30) &                        {\bf13} (15) &                       {\bf121} (10) &                       {\bf466} (10) &                        {\bf494} (7) &                     {\bf24117} (13) &                      {\bf14019} (6) &                      {\bf30631} (9) \\
                      Columba  &                        {\bf3} (382) &                       {\bf13} (255) &                       {\bf68} (144) &                      {\bf552} (146) &                      {\bf940} (113) &                     {\bf2111} (108) &                     {\bf10908} (73) &                     {\bf64606} (94) \\
                  Dnsjava   &                        {\bf6} (165) &                       {\bf13} (139) &                       {\bf101} (71) &                       {\bf218} (82) &                      {\bf1553} (54) &                      {\bf5063} (50) &                     {\bf16363} (33) &                    ${\infty}$(44)  \\
                        jEdit  &                        {\bf3} (115) &                        {\bf13} (84) &                        {\bf58} (53) &                       {\bf251} (48) &                      {\bf2906} (32) &                      {\bf3189} (30) &                      {\bf5648} (29) &                     {\bf23395} (32) \\
                        jBoss  &                        {\bf6} (514) &                       {\bf15} (353) &                       {\bf88} (208) &                      {\bf272} (189) &                     {\bf1057} (147) &                     {\bf6034} (150) &                     {\bf13148} (86) &                    {\bf38485} (113)  \\
                    jHotdraw6  &                         {\bf7} (21) &                        {\bf13} (21) &                        {\bf159} (9) &                       {\bf187} (10) &                      {\bf1779} (10) &                        {\bf611} (3) &                     ${\infty}$(5) &                      {\bf56391} (2)  \\
                        jUnit  &                         {\bf3} (40) &                        {\bf42} (39) &                       {\bf596} (18) &                    ${\infty}$(11) &                      {\bf49345} (7) &                    ${\infty}$(11) &                      {\bf31634} (9) &                     ${\infty}$(6)  \\
                    Log4j   &                        {\bf6} (223) &                       {\bf15} (134) &                       {\bf146} (68) &                       {\bf665} (70) &                      {\bf6459} (64) &                     {\bf16879} (42) &                     {\bf55582} (41) &                    ${\infty}$(48)  \\
         org.eclipse.jdt.core  &                       {\bf6} (1606) &                      {\bf26} (1025) &                       {\bf93} (657) &                      {\bf291} (631) &                     {\bf1704} (392) &                     {\bf4639} (416) &                    {\bf18344} (314) &                    {\bf74863} (309)  \\
     org.eclipse.ui.workbench  &                       {\bf3} (1184) &                       {\bf13} (783) &                       {\bf74} (414) &                      {\bf311} (464) &                     {\bf1023} (326) &                     {\bf6035} (305) &                    {\bf22864} (215) &                    {\bf77532} (192)  \\
                       Scarab  &                        {\bf6} (653) &                       {\bf16} (346) &                      {\bf113} (202) &                      {\bf420} (159) &                      {\bf764} (113) &                     {\bf3914} (137) &                     {\bf13104} (89) &                     {\bf59232} (77) \\
                   Struts   &                        {\bf3} (221) &                       {\bf17} (133) &                       {\bf100} (86) &                      {\bf222} (103) &                       {\bf675} (61) &                      {\bf4785} (77) &                     {\bf16796} (39) &                     {\bf95588} (34) \\
                   Tomcat   &                        {\bf3} (281) &                       {\bf13} (167) &                      {\bf135} (111) &                      {\bf431} (120) &                      {\bf1068} (84) &                      {\bf3497} (87) &                      {\bf7407} (61) &                     {\bf34240} (51) \\

   \hline
\end{tabular}
\end{center}
\caption{
The median number of attempts (in bold) required to find the correct repair shape of fix transactions. 
The values in brackets indicate the number of fix transactions tested per project and per transaction size for repair model CT. 
The repair model CT is made from the distribution probability of changes included in 5-SC transaction bags.
For small transactions, finding the correct repair shape in the search space is done in less than 100 attempts.
}
\label{tab:repair-ct}
\vspace{\figureskip}

\end{table*}

We have seen in Section \ref{sec:MCShaper} that the time for finding correct repair shapes depends on a probability distribution over repair actions. 
The probability distribution $\mathcal{P}$ is crucial for minimizing the search space traversal:
a good distribution P results in \emph{concentrating on likely repairs first}, i.e. the repair space is traversed in a guided way, by first exploring the parts of the space that are likely to be more fruitful. This poses two important questions: first, how to set up a probability distribution over repair actions; second, how to compare the efficiency of different probability distributions to find good repair shapes.

To compute a probability distribution over repair actions, we propose to learn them from software repositories. For instance, if many bug fixes are made of inserted method calls, the probability of applying such a repair action should be high. 
Despite our single method (learning the probability distributions from software repositories), we have shown in \ref{sec:repair-actions} that there is no single way to compute them (they depend on different heuristics). To compare different distributions against each other, we set up the following process.

One first selects bug repair transactions in the versioning history. Then, for each bug repair transaction, one extracts its repair shape (as a set of repair actions of a repair model). Then one computes the average time that a maximum likelihood approach would need to find this repair shape using equation~\ref{eq:median-repair-time}.

Let us assume two probability distributions $\mathcal{P}_1$ and $\mathcal{P}_2$ over a repair model and four fixes ($F_1\ldots F_4$) consisting of two repair actions and observed in a repository. Let us assume that the time (in number of attempts) to find the exact shape of $F_1\ldots F_4$ according to  $\mathcal{P}_1$ is
$(5, 26, 9, 12)$ and according to  $\mathcal{P}_2$ 
$(25, 137, 31, 45)$.
In this case, it's clear that the probability distribution $\mathcal{P}_1$ enables us to find the correct repair shapes faster (the shaping time for $\mathcal{P}_1$ are lower). Beyond this example, by applying the same process over real bug repairs found in a software repository, our process enables us to select the best probability distributions for a given a repair model. 

Since equation~\ref{eq:median-repair-time} is parametrized by a number of repair actions, we instantiate this process for all bug repair transactions of a certain size (in terms of AST changes). This means that our process determines the best probability distribution for a given bug fix shape size.

\subsection{Cross-Validation}

We compute different probability distributions $\mathcal{P}_x$ from  transaction bags found in repositories.
We evaluate the time to find the shape of real fixes that are also found in repositories, which may bias the results.
To overcome this problem, we use cross-validation: we always use different sets of transactions to estimate $\mathcal{P}$ and to calculate the average number of attempts required to find a correct repair shape. 
Using cross-validation reduces the risk of overfitting.

Since we have a dataset of 14 independent software repositories, we use this dataset structure for cross-validation. We take one repository for extracting repair shapes and the remaining 13 projects to calibrate the repair model (i.e. to compute the probability distributions). We repeat the process 14 times, by testing each of the 14 projects separately.
\emph{In other words, we try to predict real repair shapes found in one repository from data learned on other software projects}.

Figure \ref{bugsim} sums up this algorithm to compare fix shaping strategies.
From a bag of transactions $C$, function $split$ creates a set of testing transactions and a set of evaluation transactions. 
Then, one trains a repair model (with function $trainModel$), for repair models CT and CTET it means  computing a probability distribution on a specific bag of transactions.
Finally, for each repair of the testing data, one computes its ``repairability'' according to the repair model (with Equation \ref{eq:median-repair-time}). The algorithm returns the median repairability, i.e. the median number of attempts required to repair the test data.

\subsection{Empirical Results}
\label{sec:emp-results}

We run our fix shaping process on our dataset of 14 repositories of Java software considering two repair models: CT and CTET (see Section \ref{sec:change-models}).
We remind that CT consists of 41 repair actions and CTET of 173 repair actions.
For both repair models, we have tested the different heuristics of \ref{sec:heuristics} to compute the median repair time:  all transactions (ALL);
  one AST change (1-SC);
  5 AST changes (5-SC);
  10 AST changes (10-SC);
  20 AST changes (20-SC);
  transactions with commit text containing ``bug'', ``fix'', ``patch'' (BFP);
 a baseline of a uniform distribution over the repair model (EQP for equally-distributed probability).

We extracted all bug fix transactions with less than 8 AST changes from our dataset. For instance, the versioning repository of DNSJava contains 165 transactions of 1 repair action, 139 transactions of size 2, 71 transactions of size 3, etc.
The biggest number of available repair tests are in jdt.core (\numprint{1605} fixes consist of one AST change), while Jhotdraw has only 2 transactions of 8 AST changes.
We then computed the median number of  attempts to find the correct shape of those \numprint{23048}  fix transactions.
Since this number highly depends on the probability distributions $\mathcal{P}_x$, we computed the median repair time for all combinations of fix size transactions, projects, and heuristics discussed above ($8\times 14 \times 6)$.

\begin{figure}
\centering
\includegraphics[width=\columnwidth]{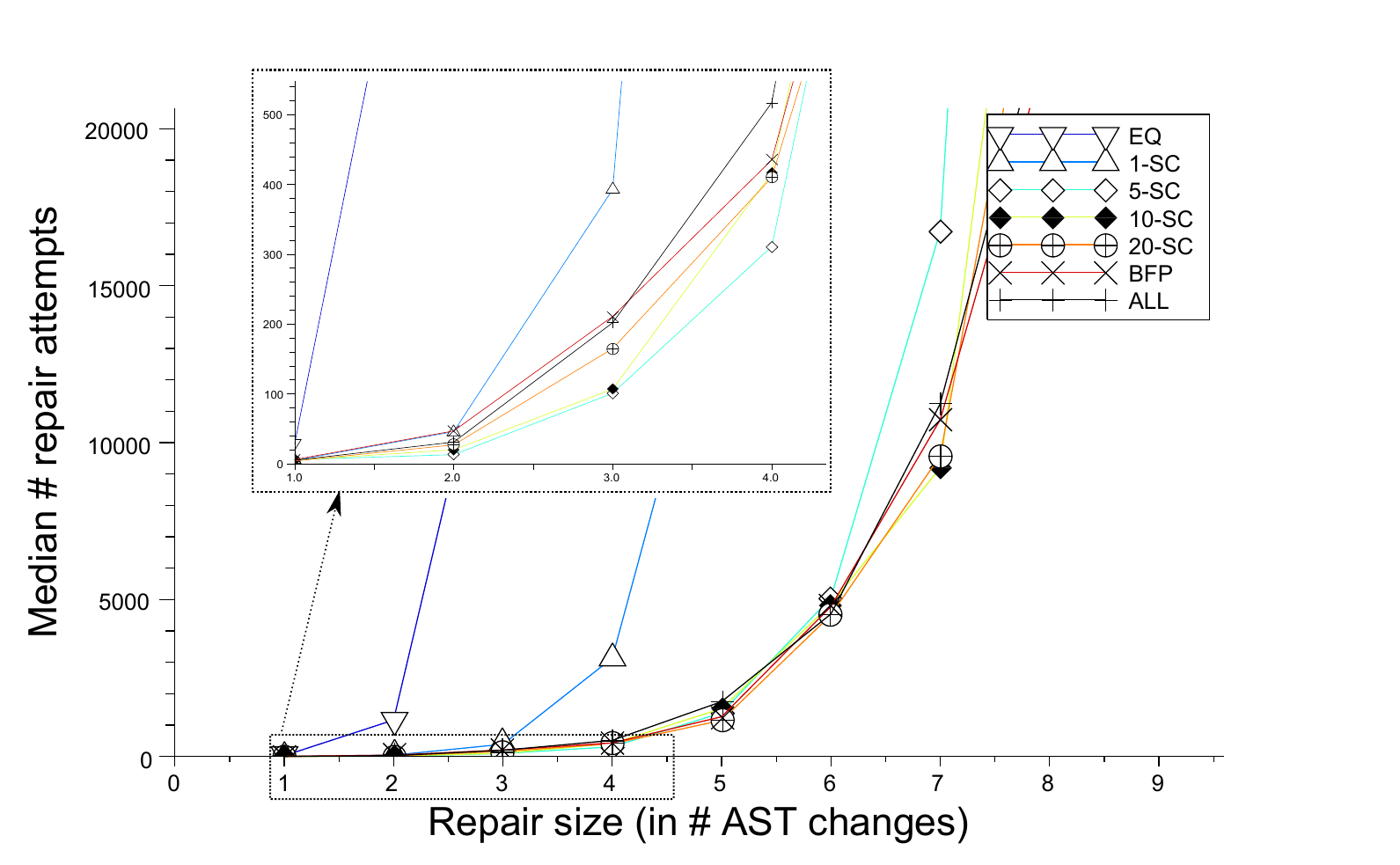}
\caption{The repairability of small transactions in repair model CT. Certain probability distributions yield a median repair time that is much lower than others.}
\label{fig:fig-repair-size-ct.pdf}
\end{figure}

Table \ref{tab:repair-ct} presents the results of this evaluation for repair space CT and transaction bag 5-SC.
For each project, the bold values give the median repairability in terms of number of attempts required  to find the correct repair shape with a maximum likelihood approach. 
Then, the bracketed values give the number of transactions per transaction size (size in number of AST changes) and per project. 
For instance, over 996 fix transactions of size 1 in the ArgoUML repository, it takes an average of 6 attempts to find the correct repair shape.
On the contrary, for the 51 transactions of size 8 in the Tomcat repository, it takes an average of \numprint{34240} attempts to find the correct repair shape.
Those results are encouraging: for small transactions, it takes a handful of attempts to find the correct repair shape. The probability distribution over the repair model seems to drive the search efficiently.
The other heuristics yield similar results -- the complete results (6 tables -- one per heuristic) are given in \cite{paperdata}.

About cross-validation, one can see that the performance over the 14 runs (one per project) is similar (all columns of Table \ref{tab:repair-ct} contain numbers that are of similar order of magnitude).
Given our cross-validation procedure, this means that for all projects, we are able to predict the correct shapes using only knowledge mined in the other projects. This gives us confidence that one could apply our approach to any new project using the probability distributions mined in our dataset.

Furthermore, finding the correct repair shapes of larger transactions (up to 8 AST changes) has an order of magnitude of $10^4$ and not more.
Theoretically, for a given fix shape of $n$ AST changes, the size of the repair model is the number of repair actions of the model at the power of $n$ (e.g. $|CT|^n$). For CT and $n=4$, this results in a space of $41^4 = \numprint{2825761}$ possible shapes (approx $10^6$). In practice, overall all projects, for small shapes (i.e. less or equal than 3 changes), a well-defined probability distribution can guide to the correct shape in a median time lower than 200 attempts. This again show that the probability distribution over the repair model is so unbalanced that the likelihood of possible shapes is concentrated on less than $10^4$ shapes (i.e. that the probability density over $|CT|^n$ is really sparse).

\begin{figure}
\centering
\includegraphics[scale=0.45]{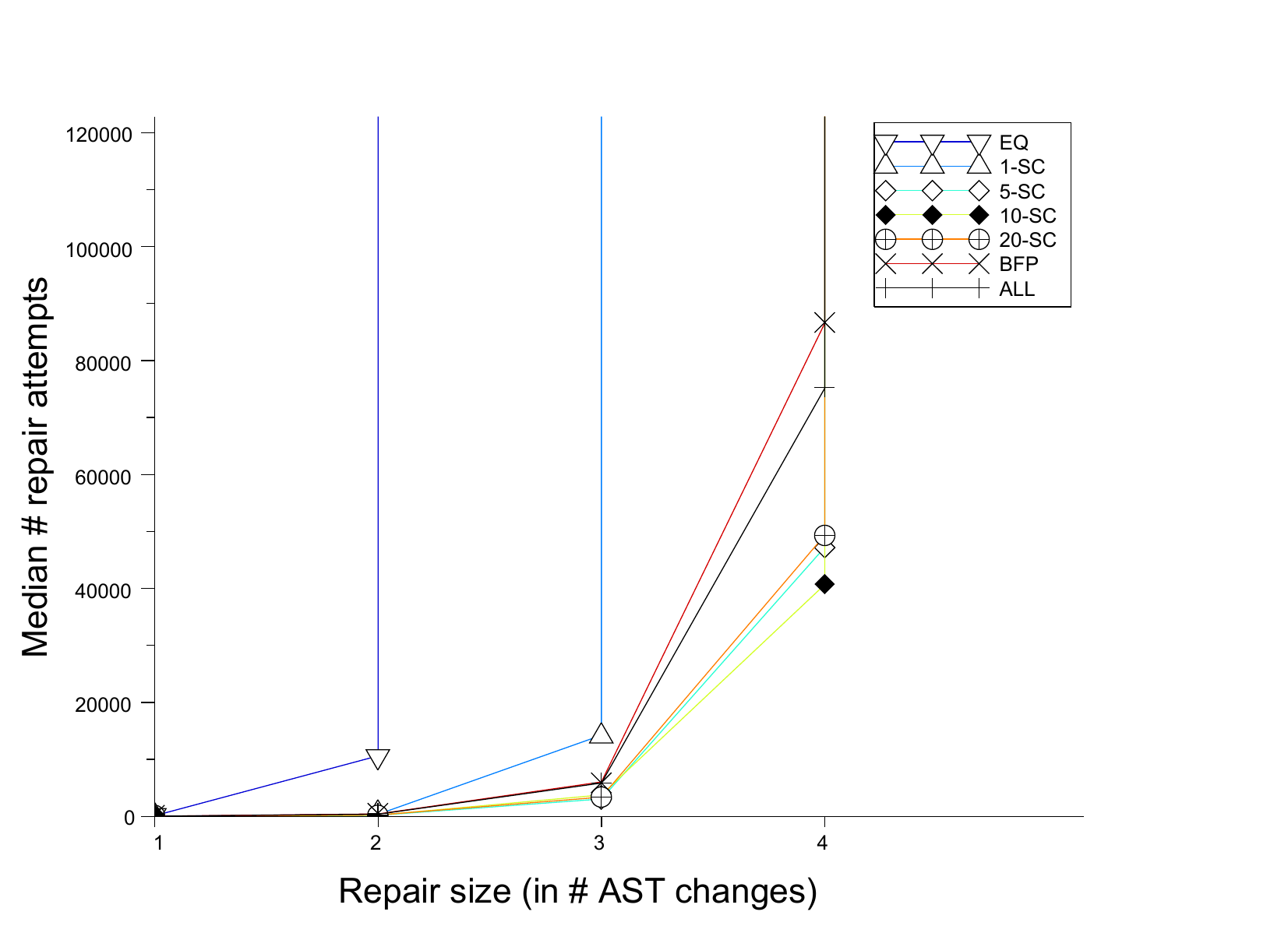}
\caption{The repairability of small transactions in repair space CTET. 
There is no way to find the repair shapes of transactions larger than 4 AST code changes.}
\label{fig:fig-repair-size-et.pdf}
\end{figure}

Now, what is the best heuristic, with respect to shaping, to train our probabilistic repair models?
For each repair shape size of Table \ref{tab:repair-ct} and heuristic, we computed the median repairability  over all projects of the dataset (a median of median number of attempts).
We also compute the median repairability for a baseline of a uniform distribution (EQP) over the repair model (i.e. $\forall i, P(r_i)=1/|CT|)$).
Figure \ref{fig:fig-repair-size-ct.pdf} presents this data for repair model CT.
It shows the median number of attempts required to identify correct repair shapes as Y-axis. The X-axis is the number of repair actions in the repair test (the size). Each line represents probability estimation heuristics.

Figure \ref{fig:fig-repair-size-ct.pdf} gives us important pieces of information.
First, the heuristics yield different repair time. 
For instance, the repair time for heuristic 1-SC is generally higher than for 20-SC.
Overall, there is a clear order between the repairability time:
for transactions with less than 5 repair actions heuristic 5-SC gives the best results, 
while for bigger transactions 20-SC is the best. 
Interestingly, certain heuristics are inappropriate for maximum-likelihood shaping of real bug fixes: the resulting distributions of probability results in a repair time that explodes even for small shape (this is the case for a uniform distribution EQP even for shape of size 3). 
Also, all median repair times tend toward infinity for shape of size larger than 9.
Finally, although 1-SC is not good over many shape size, we note that that for small shape of size 1 is better. This is explained by the empirical setup (where we also decompose transactions by shape size).

\subsubsection{On The Best Heuristics for Computing Probability Distributions over Repair Actions}

To sum up, for small repair shapes heuristic 1-SC is the best with respect to probabilistic repair shaping, but it is not efficient for shapes of size greater than two AST-level changes. 
Heuristics 5-SC and 20-SC are the best for changes of size greater than 2.
An important point is that \emph{some probability distributions (in particular built from heuristics EQP and 1-SC) are really suboptimal for quickly navigating into the search space}.

Do those findings hold for repair model CTET, which has a finer granularity?

\subsubsection{On The Difference between Repair Models CT and CTET}

We have also run the whole evaluation with the repair model CTET (see \ref{sec:change-models}). The empirical results are given in \companiondoc (in the same form as Table \ref{tab:repair-ct}).

Figure \ref{fig:fig-repair-size-et.pdf} is the sibling of figure \ref{fig:fig-repair-size-ct.pdf} for repair model CTET. 
They look rather different.
The main striking point is that with repair model CTET, we are able to find the correct repair shape for fixes that are no larger than 4 AST changes. 
After that, the arithmetic of very low probabilities results in virtually infinite time to find the correct repair shape. On the contrary, in the repair model CT, even for fixes of 7 changes, one could find the correct shape in a finite number of attempts.
Finally, in this repair model the average time to find a correct repair shape is several times larger than in CT (in CT, the shape of fixes of size 3 can be find in approx. 200 attempts, in CTET, it's more around \numprint{6000}).

For a given repair shape, the synthesis consists of finding concrete instances of repair actions.
For instance, if the predicted repair action in CTET consists of inserting a method call, it remains to predict the target object, the method and its parameters.
We can assume that the more precise the repair action, the smaller the ``synthesis space''.
For instance, in CTET, the synthesis space is smaller compared to CT, because it only composed of enriched versions of basic repair actions of repair model CT (for instance inserting an ``if'' instead of inserting a statement).

Our results illustrate the tension between the richness of the repair model and the ease of fixing bugs automatically.
When we consider CT, we find likely repair shapes quickly (less than 5,000 attempts), even for large repair, but to the price of a larger synthesis space.
In other words, there is a balance between finding correct repair actions and finding concrete repair actions. 
When the repair actions are more abstract, it results in a larger synthesis space, when repair actions are more concrete, it hampers the likelihood of being able to concentrate on likely repair shapes first.
We conjecture that the profile based on CT is better because of the following two points:  it enables us to find bigger correct repair shapes (good) in a smaller amount of time (good). 

Finally, we think that our results empirically explore some of the foundations of ``repairing'': there is a difference between prescribing aspirin (it has a high likelihood to contribute to healing, but only partially) and prescribing a specific medicine (one can try many medicines before finding the perfect one).

\section{Actionable Guidelines for Automated Software Repair}
\label{sec:guidelines}

Our results blend empirical findings with theoretical insights.
How can they be used within a approach for automated software repair?
This section presents actionable guidelines arising from our results.
We apply those guidelines in a case study that consists of reasoning on a simplified version of GenProg within our probabilistic framework.

\subsection{Consider Using a Probability Distribution over Repair Actions}
\label{guidelineA}

Automated software repair embed a set of repair actions, either explicitly or implicitly.
On two different repair models, we have shown that the importance of each repair action greatly varies. Furthermore, our mathematical analysis has proved that considering a uniform distribution over repair actions is extremely suboptimal. 

Hence, from the viewpoint of the time to fix a bug, we recommend to set up a probability distribution over the considered repair actions.
This probability distribution can be learned on past data as we do in this paper or simply tuned with an incremental evaluation process. 
For instance, Le Goues et al. \cite{GouesWF12} have done similar probabilistic tuning over their three repair actions.
\emph{Overall, using a probability distribution over repair actions could significantly fasten the repair process.}

\subsection{Be Aware of the Interplay between Shaping and Synthesis}

We have shown that having more precise shapes has a real impact on shaping time. 
In repair model CT, for fix shapes of size 3, the logical shaping time is approximately 150 attempts.
In repair model CTET, for shapes of the same size, the average logical time jumps around \numprint{4000}, which represents more than a ten-fold increase.
Our work quantitatively highlights the impact of consider more precise repair actions.
\emph{By being aware of the interplay between shaping and synthesis, the research community will be able to create a disciplined catalog of repair actions and to identify where the biggest synthesis challenges lie.}

\subsection{Analyze the Repairability depending on The Fix Size}

We have shown that certain repair shapes are impossible to find because of their size. In repair model CT, the shapes of more than 10 repair actions are not found in a finite time. In repair model CTET, the repair shapes of more than 5 actions are not found either. Given that a repair shape is an abstraction over a concrete bug fix, if one can not find the abstraction, there is no chance to find the concrete bug fix.

Our analysis for identifying this limit is agnostic of the repair actions. Hence one can use our methodology and equation to analyze the size of the ``findable'' fixes. \emph{Our probabilistic framework enables one to understand the theoretical limits of certain repair processes.} 

Let us now apply those three guidelines on a small case study.

\subsection{Case Study: Reasoning on GenProg within our Probabilistic Framework}
\label{ref:gcd}

\begin{lstlisting}[label={gcd},caption={The infinite loop bug of Weimer et al's bug \cite{Weimer2009}. Code insertion can be made on 13 places, 8 AST subtrees can be deleted or copied.},firstline=7,lastline=33,float]
    // insert 1
    if (a == 0) { // ast 1
      // insert 2
      System.out.println(b); // ast 2
      // insert 3
    }
    // insert 4
    while (b != 0) { // infinite loop // ast 3
      // insert 5
      if (a > b) { // ast 4
        // insert 6
        a = a - b; // ast 5
        // insert 7
      } else {
        // insert 8
        b = b - a; // ast 6
        // insert 9
      }
      // insert 10
    }
    // insert 11
    System.out.println(a); // ast 7
    // insert 12
    return; // ast 8
    // insert 13
  }
\end{lstlisting}

We now aim at showing than our model also enables to reason  on Weimer et al's \cite{Weimer2009} example program. This program, shown in Listing \ref{gcd},  implements Euclid's greatest common divisor algorithm, but runs in an infinite loop if $a=0$ and $b>0$.
The fix consists of adding a ``return'' statement on line 6.

\paragraph{Probability Distribution}
In Weimer et al's repair approach, the repair model consists of three repair actions: inserting statements, deleting statements, and swapping statements\footnote{In more recent versions of GenProg, swapping has been replaced by ``replacing''.}. By statements, they mean AST subtrees. 
With a uniform probability distribution, the logical time to find the correct shape is 4 (from Equation \ref{eq:median-repair-time}). If one favors insertion over deletion and swap, for instance by setting $p_{insert=0.6}$, the median logical time to find the correct repair action becomes 2 which is twice faster. Between 2 and 4, it seems negligible, but for larger repair models, the difference might be counted in days, as we show now.

\paragraph{Shaping and Synthesis}
In the GCD program, there are $n_{place} = 13$ places where $n_{ast} = 8$ AST statements can be inserted. 
In this case, the size synthesis space can be formally approximate: 
the number of possible insertion is $n_{place}*n_{ast}$;
the number of possible deletion is $n_{ast}$;
the number of possible swap is $(n_{ast})^2$.

This enables us to apply our probabilistic reasoning at the level of concrete fix as follows.
We define the concrete repair distribution as:
$
p_{insert}(ast_i, place_k) = \frac{p_{insert}}{n_{place}*n_{ast}}
$,
$
p_{delete}(ast_j) = \frac{p_{delete}}{n_{ast}}
$,
$
p_{swap}(ast_i, ast_j) = \frac{p_{insert}}{(n_{ast})^2}
$.

With a uniform distribution $p_{insert} = p_{delete} =  p_{swap} = 1/3$, formula \ref{eq:median-repair-time} yields that the logical time to fix this particular bug (insertion of node \#8 at place \#3) is 219 attempts (not that it is not anymore a shaping time, but the real number of required runs).
However, we observed over real bug fix that $p_{insert} > p_{delete}$ (see Table \ref{tab:top-10-ct-heuristics}).
What if we distort the uniform distribution over the repair model to favor insertion? The following table gives the results for arbitrary distributions spanning different kinds of distribution:

\begin{center}
\begin{tabular}{c|c|c|c}
$p_{insert}$ & $p_{delete}$ & $p_{swap}$ & Logical time \\
\hline
.33& .33& .33 & 219\\
.39& .28& .33 & 185\\
.45& .22& .33 & 160\\
.40& .40& .20 & 180\\
.50& .30& .20 & 144\\
.60& .20& .20 & 120\\
\end{tabular}
\end{center}

This table shows that as soon as we favor insertion over deletion of code, the logical time to find the repair do actually decrease.  

Interestingly, the same kind of reasoning  applies to fault localization. Let's assume that a fault localizer filters out half of the possible places where to modify code (i.e. $n_{place} = 7$). Under the uniform distribution and the space concrete repair space, the logical time to find the fix decreases from 219 to 118 runs.

\paragraph{Repairability and Fix Size}
We consider the same model but on larger programs with fault localization, for instance 100 AST nodes and 20 potential places for changes.
Let us assume that the concrete fix consists of inserting node \#33 at place \#13.
Under a uniform distribution, the corresponding repair time according to formula \ref{eq:median-repair-time} is $\geq \numprint{20000}$ runs.
Let us assume that the concrete fix consists of two repair actions:
inserting node \#33 at place \#13 and  deleting node \#12.
Under a uniform distribution, the repair time becomes \numprint{636000} runs, a 30-fold increase.

Obviously, for sake of static typing and runtime semantics, the nodes can not be inserted anywhere, resulting in lower number of runs. However, we think that more than the logical time, what matters is the order of magnitude of the difference between the two scenarios.
Our results indicate that it is very hard to find concrete fixes that combine different repair actions.

Let us now be rather speculative. Those simulation results contribute to the debate on whether past results on  evolutionary repair are either evolutionary or guided random search \cite{Arcuri2011}. According to our simulation results, it seems that the evolutionary part (combining different repair actions) is indeed extremely challenging. On the other hand, our simulation does not involve fitness functions, it is only guided random search, what we would call ``Monte Carlo'' repair. A good fitness function might counter-balance the combinatorial explosion of repair actions.

\section{Related Work}
\label{sec:relatedwork}

\paragraph{Empirical Studies of Versioning Transactions}

Purushothaman and Perry \cite{Purushothaman2005} studied small commits (in terms of number of lines of code) of proprietary software at Lucent Technology. They showed the impact of small commits with respect to introducing new bugs, and whether they are oriented toward corrective, perfective or adaptive maintenance.
German \cite{German2006} asked different research questions on what he calls ``modification requests'' (small improvements or bug fix), in particular with respect to authorship and change coupling (files that are often changed together).
Alali and colleagues \cite{Alali2008} discussed the relations between different size metrics for commits (\# of files, LOC and \# of hunks), along the same line as Hattori and Lanza \cite{Hattori2008} who also consider the relationship between commit keywords and engineering activities.
Finally, Hindle et al. \cite{Hindle2008,Hindle2009} focus on large commits, to determine whether they reflect specific engineering activities such as license modifications.
Compared to these studies on commits that mostly focus, on metadata (e.g. authorship, commit text) or size metrics (number of changer files, number of hunks, etc.), we discuss the content of commits and the kind of source code change they contain.
Fluri et al. \cite{Fluri2008} and Vaucher et al. \cite{Vaucher2008} studied the versioning history to find patterns of change, i.e. groups of similar versioning transactions. 

Pan et al. \cite{Pan2008} manually identified 27 bug fix patterns on Java software. Those patterns are precise enough to be automatically extractable from software repositories. They provide and discuss the frequencies of the occurrence of those patterns in 7 open source projects. This work is closely related to ours: we both identify automatically extractable repair actions of software. The main difference is that our repair actions are discovered fully automatically based on AST differencing (there is no prior manual analysis to find them). Furthermore, since our repair actions are meant to be used in an automated program repair setup, they are smaller and more atomic. 

Kim and et al. \cite{Kim2006a} use versioning history to mine project-specific bug fix patterns.
Williams and Hollingsworth \cite{Williams2005} also learn some repair knowledge from versioning history. They mine how to statically recognize where checks on return values should be inserted. Livshits and Zimmermann \cite{Livshits2005} mine co-changed method calls. 
The difference with those close pieces of research is that we enlarge the scope of mined knowledge:
from project-specific knowledge \cite{Kim2006a} to domain-independant repair actions, and from one single repair action \cite{Williams2005,Livshits2005} to 41 and 173 repair actions.

\paragraph{Abstract Syntax Tree Differencing}

The evaluation of AST differencing tools often gives hints about common change actions of software. For instance, Raghavan et al. \cite{Raghavan2004} showed the six most common types of changes for the Apache web server and the GCC compiler, the number one being ``Altering existing function bodies''. This example clearly shows the difference with our work: we provide change and repair actions at a very fine granularity.
Similarly, Neamtiu et al. \cite{Neamtiu2005} gives interesting numerical findings about software evolution such as the evolution of added functions and global variables of C code. It also remains at granularity that is coarser compared to our analysis.
Fluri et al. \cite{Fluri2007b} gives some frequency numbers of their change types in order to validate the accuracy and the runtime performance of their distilling algorithm. Those numbers were not --- and not meant to be --- representative of the overall abundance of change types. Giger et al. \cite{Giger2012} discuss the relations between 7 categories of change types and not the detailed change actions as we do.

\paragraph{Automated Software Repair}

We have already mentioned many pieces of work on automated software repair (incl. \cite{weimer2006patches,Weimer2009,Wei2010,Dallmeier2009,Arcuri20113494,Carzaniga2010}).
We have discussed in details the relationship of our work with GenProg.
Let us now compare with the other close papers.

Wei et al. \cite{Wei2010} presented AutoFix-E, an automated repair tool which works with contracts. In our perspective, AutoFix-E is based on two repair actions: adding sequences of state-changing statements (called ``mutators'') and adding a precondition (of the form of an ``if'' conditional). Their fix schemas are combinations of those two elementary repair actions. In contrast, we have 173 basic repair actions and we are able to predict repair shapes that consist of combinations of 4 repair actions. However, our approach is more theoretical than theirs.
Our probabilistic view on repair may fasten their repair approach: it is likely that not all ``fix schemas'' are equivalent. For instance, according to our experience, adding a precondition is a very common kind of fix in real bugs.

Debroy et al. \cite{debroy2010using} invented an approach to repair bugs using mutations inspired from the field of mutation testing.
The approach uses a fault localization technique to obtain  the candidate faulty locations. For a given location, it applies mutations, producing mutants of the program. Eventually, a mutant is classified as ``fixed'' if it passes the test suite of the program.
Their repair actions are composed of mutations of arithmetic, relational, logical, and assignment operators.
Compared to our work, mutating a program is a special kind of fix synthesis where no explicit high-level repair shapes are manipulated.
Also, in the light of our results, we assume that a mutation-based repair process would be faster using probabilities on top of the mutation operators.

Kim et al. \cite{KimPAR2013} introduced \emph{PAR}, an algorithm that generates program patches using a set of 10 manually written fix templates. 
As GenProg, the approach leverages evolutionary computing techniques to generate program patches. 
We share with PAR the idea of extracting repair knowledge from human-written patches.
Beyond this high-level point in common, there are three important differences.
First, they do a manual extraction of fix patterns (by reading \numprint{62656} patches) while we automatically mine them from the past commits. 
Second, PAR patterns and our repair actions are expressed at a different granularity. 
PAR patterns contain a specification of the context that matches a piece of AST, a specification of analysis (e.g. to collect compatible expressions in the current scope), and a specification of change. Our repair actions correspond to this last part. While their patterns are operational, their change specifications are ad hoc (due to the process of manually specifying templates). On the contrary, our specification of repair actions are systematic and automatically extracted, but our approach is more theoretical and we do not fix concrete bugs.
This shows again that the foundations of their approach contains more manual work than ours: a PAR pattern is a \emph{manually} identified repair schema where all the synthesis rules are \emph{manually} encoded.
Finally, we think it is possible to marry our approaches by decorating their templates with probability distributions (whether mined or not) so as to speed up the repair.

\section{Conclusion}

In this paper, we have presented the idea that one can mine repair actions from software repositories.
In other words, one can learn from past bug fixes the main repair actions (e.g. adding a method call). Those repair actions are meant to be generic enough to be independent of the kinds of bug and the software domains.
We have discussed and applied a methodology to mine the repair actions of \numprint{62179} versioning transactions extracted from 14 repositories of 14 open-source projects.
We have largely discussed the rationales and consequences of adding a probability distribution on top of a repair model. We have shown   that certain  distributions over repair actions can result in an infinite time (in average) to find a repair shape while other fine-tuned distributions enable us to find a repair shape in hundreds of repair attempts.

The main direction of future work consists of going beyond empirical results and theoretical analysis.
We are now exploring how to use this learned knowledge (of the form of probabilistic repair models) to fix real bugs.
In particular, we are planning to work on using probabilistic models to see whether one can faster repair the bugs of PAR's and GenProg's datasets.
The latter involves having a Java implementation of GenProg and would advance our knowledge on whether GenProg's efficiency is really language-independent (Segfaults and buffer overruns do not exists in Java).

\bibliographystyle{ieeetr}   
\balance
\bibliography{biblio-software-repair}

\end{document}